# Artifact Correction in Magnetic Resonance Temperature Imaging for Laser Interstitial Thermotherapy with Multi-echo Acquisitions

Ziyi Pan[1#], Yuancheng Jiang[1#], Wenbo Lv[1], Sisi Li[1], Meng Han[2], Yawei Kuang[2], Hao Sun[2], Xiu Wang[3], Jianjun Bai[4], Wenbo Liu[2], Guangzhi Wang[5], and Hua Guo[1*]

[1]Center for Biomedical Imaging Research, School of Biomedical Engineering, Tsinghua University, Beijing, China

[2]Sinovation Medical, Beijing, China

[3]Department of Neurosurgery, Beijing Tiantan Hospital, Capital Medical University, Beijing, China

[4]Tsinghua University Yuquan Hospital, Beijing, China

[5]School of Biomedical Engineering, Tsinghua University, Beijing, China

# Ziyi Pan and Yuancheng Jiang contributed equally to this work.

**Running title:** Signal Void Correction for MRgLITT

**Word Count (body):** ~6000

**\*Correspondence to:**

Hua Guo, Ph.D.

Center for Biomedical Imaging Research

School of Biomedical Engineering, Tsinghua University, Beijing, China

Phone: +86-010-6279-5886

Email: huaguo@tsinghua.edu.cn

# Abstract

**Background:** In MRI-guided laser interstitial thermotherapy (MRgLITT), a signal void sometimes appears at the heating center of the measured temperature map. This can greatly affect the accuracy of temperature mapping and increase the risk of over- or undertreatment. In neurosurgical MRgLITT treatments, cerebrospinal fluid pulsation (CSF), which may lead to temperature artifacts, also needs to be carefully managed. Additionally, accurate temperature imaging requires a robust phase unwrapping procedure that is resistant to artifacts.

**Purpose:** We find that signal loss in MR magnitude images can be one distinct contributor to the temperature imaging signal void. Therefore, this study aims to investigate this finding and more importantly, to correct the signal loss-induced temperature errors for reliable temperature imaging. Also, this study intends to improve measurement accuracy by correcting CSF-induced temperature errors and employing a more reliable phase unwrapping algorithm.

**Methods:** First, a gradient echo (GRE) sequence with certain TE values for temperature imaging is used to quantify T2* variations during MRgLITT heating and to investigate the development of signal voids throughout the treatment. Then, informed by these findings, a multi-echo GRE sequence with appropriate TE coverage is employed. A multi-echo-based correction algorithm is developed to address the signal loss-induced temperature errors by utilizing short echo times (e.g., TE = 4 ms) that are less affected by the signal loss. Additionally, a new phase unwrapping method and a new CSF pulsation correction approach are developed for multi-echo signal processing. Finally, the temperature imaging method is evaluated by gel phantom, ex-vivo (pork and swine brain), and in-vivo (canine brain) LITT heating experiments.

**Results:** T2* shortening during heating can be one important cause of the temperate imaging signal voids and this demands the multi-echo acquisition with varied TE values. Ex-vivo and in-vivo experiments demonstrate that the proposed multi-echo-based method can effectively correct signal loss-induced temperature errors. Compared with conventional single-echo thermometry, the multi-echo thermometry showed lower root-mean-square errors (RMSE = 0.175 ℃ for gel phantom and 0.423 ℃ for ex-vivo). According to neurosurgeons' ratings, the multi-echo thermometry in the in-vivo experiments shows smoother hotspot boundaries, fewer artifacts in temperature maps, and improved thermometry reliability compared to traditional single-echo thermometry. In the in-vivo experiments, the ablation areas estimated from the multi-echo thermometry also show satisfactory agreement with those determined from post-ablation MR imaging.

**Conclusions:** Temperature imaging signal voids can be compensated using the multi-echo GRE sequence together with the multi-echo-based correction algorithm.

# Introduction

MRI-guided laser interstitial thermotherapy (MRgLITT) is a minimally invasive therapy that serves as an alternative option for treating lesions in surgically challenging locations, whether anatomically or functionally[1-4]. Heating in the range of 44 ~ 90℃ (with 90℃ typically set as the high-temperature limit near the laser applicator tip to prevent vaporization[5]) is applied for tens of seconds to rapidly coagulate tissue and induce cell necrosis through protein denaturation[6]. Compared to open surgery, MRgLITT offers more precise targeting of lesions, reduced risk of infections, and shorter hospital stays for patients[5,7-9].

During LITT treatment, magnetic resonance temperature imaging (MRTI) plays a crucial role in accurately ablating lesions while preserving surrounding healthy tissues. Currently, the most widely used MRTI technique for MRgLITT is proton resonance frequency (PRF) shift-based thermometry[10-13] owing to its high linearity and sensitivity for measurement across a wide temperature range (-15 ℃ ~ 100 ℃)[10]. Theoretically, the PRF shift-based method relies on the temperature dependence of hydrogen bonding in water protons. The temperature change alters the local magnetic field and is thus monitored by phase mapping using gradient echo (GRE) type sequences.

A major concern for PRF shift-based thermometry is that the accuracy of temperature measurement can be influenced by various factors. On the one hand, physical factors such as $B_0$ field drift[14], magnetic susceptibility[15], and motion[16,17] can cause considerable impacts on temperature imaging. Specifically, during the MRgLITT procedures within the brain, cerebrospinal fluid (CSF) pulsations can contribute to signal fluctuations of GRE magnitudes. These further impact phase mapping[18,19] and subsequently decrease the accuracy of temperature measurements. On the other hand, from the data processing perspective, phase wrapping[20] poses a significant challenge especially when large temperature variations occur during ablation. The presence of other issues, such as local $B_0$ field inhomogeneity, can further complicate the phase unwrapping process. Consequently, the errors of phase unwrapping result in temperature mapping artifacts and increase the risk of over- or undertreatment. The solutions to the aforementioned challenges have been explored in previous studies to improve the accuracy of temperature measurement. For instance, multi-baseline thermometry[16,21] and referenceless thermometry[17,22,23] are proposed to solve the motion-induced errors. Meanwhile, the referenceless approach also provides effective $B_0$ drift correction. Model-based methods[24,25] are suggested to correct the heat-induced susceptibility and chemical shift artifacts, especially in high fat-content tissues[26,27]. Furthermore, to avoid phase wrapping, complex phase subtraction between successive dynamic images[28,29] is employed. Besides, various phase unwrapping algorithms[30,31] have been proposed to address the phase wrapping issue.

Despite these improvements, several problems remain to be investigated in MRgLITT. One crucial issue is the common presence of black holes at the heating center on temperature maps [32,33], which is referred to as "MRTI signal void" in this study. The signal void can significantly impact the accuracy of temperature mapping and increase the risk of ablation over- or undertreatment. For example, without accurate temperature monitoring at the heating center, tissue vaporization and carbonization, which occur above 100-110℃, may take place and limit the effectiveness of laser ablation[34,35]. Furthermore, previous studies[32,33] also showed that the MRTI signal void may lead to a larger discrepancy between the thermal damage estimated by intraoperative MRTI and the ablation area measured by postoperative MRI[32,33]. Partly owing to this signal void, temperature misestimates are sometimes inevitable and likely contribute to LITT overheating. Reported complications caused by LITT overheating include new or worsening neurological deficits such as dysphasia[3], paresis or paraplegia[7,8], homonymous hemianopia[36], and short-term memory loss[37]. However, the cause of the MRTI signal void is still unclear. Therefore, further investigation and correction of the MRTI signal void are of great clinical importance.

To improve the precision and accuracy of temperature measurement, the TE values must also be carefully chosen. In the PRF shift-based temperature measurement, the signal-to-noise ratio (SNR) of the temperature change-induced phase difference ($SNR_{\Delta\phi}$) is proportional to $TE \cdot \exp(-TE/T_2^*)$ [28]. Thus an optimal TE = T2* gives the maximal $SNR_{\Delta\phi}$, and provides the best temperature-to-noise ratio (TNR[28]) and the highest temperature precision. However, in practice, TE is chosen as a trade-off between high precision (using longer TE) and high acquisition efficiency (using shorter TE and TR for temporal resolution). As such, TE around 20 ms is typically applied for single-echo GRE acquisition in clinical MRgLITT treatments at 3T[2,8]. Considering TE-related temperature mapping artifacts, imaging with shorter TE values is susceptible to the influence of motion, such as the CSF pulsation. This is because the temperature change-induced phase difference is small at these TE values, which consequently amplifies the effect of the motion-induced phase. However, short TE is helpful to avoid signal voids in MRTI. In the following sections, it will be mentioned that the accuracy of temperature measurement is significantly impacted by the considerable signal voids in MRTI when long TEs are used (e.g., ~ 20 ms).

In the past decade, the multi-echo GRE acquisition for PRF thermometry has been widely applied. One major benefit of multi-echo thermometry is that it can increase the readout bandwidth while keeping the total readout duration[38], resulting in reduced vulnerability to off-resonance with a maintained TNR[39,40]. With this benefit, multi-echo GRE acquisitions are employed to enhance the MRTI precision in MRgLITT[41]. Furthermore, by utilizing the inherent properties of short TE images, multi-echo imaging also holds the potential for signal loss compensation. For instance, a dual-echo GRE sequence with z-shimming at the first echo was reported to partly recover central magnetic susceptibility-induced signal loss near a metallic ablation probe in microwave heating[42].

In this study, the underlying cause of the previously reported MRgLITT signal void is explored. We suppose that the MRTI signal void can be attributed to the signal loss of GRE magnitude images, of which one important factor is T2* shortening at high temperatures. To validate this, T2* variations during heating are quantified using multi-echo GRE data and temperature imaging with certain TE values is performed to investigate how MRTI signal voids occur during LITT treatment. Then based on the analysis of the MRTI signal voids, a multi-echo-based GRE acquisition protocol with suitable TEs is adopted and a reliable data processing pipeline is proposed to reduce the MRTI signal voids. Additionally, in the multi-echo thermometry, a robust multi-echo-based phase unwrapping method is presented to process the severe phase wraparounds caused by high temperatures. A motion compensation method is also integrated to correct TE-independent phase variations from CSF pulsation. The multi-echo thermometry is evaluated using gel phantom, ex-vivo pork and swine brain, and in-vivo canine brain heating experiments. The multi-echo thermometry is also compared with conventional single-echo thermometry. The results demonstrate that magnitude signal loss is a significant source of temperature mapping errors, and the artifacts can be minimized by the proposed multi-echo-based method.

# Methods

## *2.1 Evaluation of Temperature Imaging Signal Voids*

### 2.1.1 Signal Voids Analysis

To understand the formation of the MRTI signal void, an initial experiment was conducted to assess the single-echo GRE image magnitude and phase variations during laser heating. A set of temperature data from in-vivo canine MRgLITT ablation (Canine 01) was utilized for this purpose. Detailed information about this experiment can be found in the 2.4.2 subsection below. A single-echo GRE sequence (i.e., TE = 19 ms) was employed, which is typically applied in clinical settings. For analysis, four separate voxels were evaluated: one voxel within the heating zone affected by the MRTI signal void, one pixel within the heating zone unaffected by the signal void, one pixel outside the heating zone in the canine brain, and one pixel in the background noise.

### 2.1.2 T2* Calculation

To better understand and correct the temperature imaging signal void, the T2* values were measured from this set of in-vivo canine MRgLITT data. A multi-echo GRE sequence (i.e., TE = 4/9/14/19 ms) was utilized for the T2* mapping. Assuming a mono-exponential decay of T2*, the signal magnitude, including the presence of noise, can be described as follows[40]:

$$S_{TEi} = S_0 \cdot \exp(-\frac{TEi}{T_2^*}) + \delta, \qquad [1]$$

where $S_{TEi}$ is the signal magnitude at the *ith* TE, and $\delta$ represents a non-zero baseline taking into account the effective noise variance. T2* maps were computed by MATLAB (Mathworks, Natick, MA, USA) open-source software qMRLab[43] (mono_t2 model[44]) for each time point.

## *2.2 Temperature Imaging Using Different TE Values*

Based on the T2* variations obtained in the 2.1.2 subsection, a multi-echo acquisition with an appropriate TE range is necessary because, as will be discussed in the *Results* section, the T2* values vary both spatially and temporally. When T2* values decrease, a shorter TE needs to be employed so as to prevent the T2* decay-induced signal loss and to increase the TNR since TE = T2* offers the theoretically best TNR. On the other hand, temperature data with relatively longer TEs are also required for time points when T2* values are large. Consequently, combining the multiple TEs will result in improved temperature measurement accuracy.

Before proposing the multi-echo-based temperature data processing algorithm described in the following subsection, an experiment was first conducted using different TE settings to assess the temperature imaging quality related to TE, particularly regarding the occurrence of MRTI signal voids. Because the measured T2* values in the in-vivo canine experiment could be shortened to less than 10 ms (will be mentioned in the *Results* below), TEs of ~10 ms and shorter (~5 ms) were employed for short T2*, while TEs of ~15 ms and longer (~ 20 ms, a typical TE value in clinic) were applied for long T2* values (~30-40 ms). A set of data from the ex-vivo swine brain (TE = 5.5/10/14.5/19 ms, continuous heating) and a set of data from the in-vivo canine brain (TE = 4/9/14/19 ms, Canine 01) were utilized. More details on these experiments were provided in the 2.4.2 subsection. For each of the four TE values, the conventional single-echo thermometry (described in the 2.3.2 subsection) was carried out. Then the magnitude images, phase images, and temperature maps derived from the four TE values were carefully compared and assessed.

## *2.3 Temperature Data Processing Algorithm*

### 2.3.1 Multi-echo Thermometry for Signal Void Correction

A multi-echo-based temperature data processing algorithm (**Figure 1**) is developed to correct the MRTI signal voids. The method enhances temperature measurement accuracy by leveraging the advantages of both short and long TE signals. The algorithm is implemented via MATLAB. The term MEC-Corr (i.e., multi-echo combined, artifacts corrected MRTI) will be used in the following to refer to the temperature measurements calculated from the entire processing pipeline. The details of each step are described below.

### Step 1. Phase Difference Map Calculation

First, from the GRE image series, the five frames (i.e., Frame 2-6) acquired before heating are averaged to provide a phase-reference map for each TE. The first frame (i.e., Frame 1) is discarded to avoid unstable signals. Then a complex-phase subtraction procedure[45] is performed to calculate the phase difference map between the current frame (**Figure S1a**) and the reference map for each TE. However, substantial temperature variations can still cause phase wraparounds (**Figure S1b**, white arrow), especially for long TEs. Such a wraparound can be avoided by employing a running complex subtraction[28,29] between successive dynamic images. However, this approach is not applicable here, because it accumulates phase errors once the signal loss-induced errors appear (which will be further illustrated in the Results section). Thus, we utilize a modified phase unwrapping algorithm as described in Step 2.

**Step 2. Multi-echo-based Phase Unwrapping for Large Temperature Changes**

A dedicated phase unwrapping method is proposed to unwrap the phase discontinuities that occur at high temperatures (**Figure S1b**, white arrow) for the multi-echo dynamic images. In practice, the shortest TE (~4 ms) used in this study is short enough such that the phase difference map $\Delta\phi_{TE1}$ is not wrapped by the large temperature changes or corrupted by signal loss within the temperature range of interest during LITT heating. Therefore, $\Delta\phi_{TE1}$ can be used to estimate the unwrapped phase difference maps $\Delta\phi_{TEi\ estimated}$ for the left three TEs (if a 4-echo GRE is used) using the following equation:

$$\Delta\phi(x,y)_{TEi\ estimated} = \frac{\Delta\phi(x,y)_{TE1}}{TE1} \times TEi,\ i = 2,3,4.$$ [2]

According to Equation [2], the unwrapped phase difference map for the left three TEs can be obtained by finding $k \in Z$ for which the difference between the unwrapped phase difference $\Delta\phi_{TEi\ unwrapped} = \Delta\phi_{TEi} + 2\pi \cdot k$ and the estimated phase difference $\Delta\phi_{TEi\ estimated}$ is smallest. Even for a large TE (such as TE4 = 19 ms), a phase change $2\pi$ corresponds to a large temperature shift (around 41 ℃), so the estimation based on TE1 can always fall within a reliable range.

**Step 3. B₀ Drift Correction**

Before LITT heating, four thermally isolated ROIs (**Figure S1a**) are manually prescribed to rectify the B₀ phase drifts induced by system instability. For each phase difference map at a particular time, the phase drift is compensated for by subtracting the mean phase difference of the four isolated ROIs. Then, using the PRF shift-based approach, temperature images (**Figure S1d**) for the four TEs can be derived from the unwrapped and B₀ drift-corrected phase difference maps (**Figure S1c**) using Equation [3].

$$\Delta T_{TEi} = \frac{\Delta\phi_{TEi}}{\alpha\gamma B_0 TEi}\ ,$$ [3]

where $\Delta\phi_{TEi}$ is the phase difference between the current image phase $\phi(T)$ and the reference baseline phase $\phi(T_0)$ at *TEi,* $\alpha$ is the PRF change coefficient with a value of -0.01 ppm / ℃, $\gamma$ is the gyromagnetic ratio, and B₀ is the static field strength.

**Step 4. Correction of the Signal Loss-Induced Error**

In Step 4, the first echo (**Figure S1e**) is utilized to compensate for the signal loss-induced errors for the left TEs whose temperature images are corrupted by the signal void. Specifically, $\Delta T(x,y)_{TEi}$ ($i = 2,3,4$) for each pixel within the heating center is examined. If $| \Delta T(x,y)_{TE1} - \Delta T(x,y)_{TEi}|$ exceeds a pre-set threshold of 5 ℃ and the normalized magnitude $S(x,y)_{TEi}$ is below a threshold of 0.2, we consider the signal loss corrupts the temperature measurement and thus replace $\Delta T(x,y)_{TEi}$ by $\Delta T(x,y)_{TE1}$. The thresholds are empirically set. The signal loss detection and correction are further illustrated in **Figure S1e - g**.

**Step 5. Correction of the CSF-Induced Error**

Step 5 is to reduce the influence of CSF pulsation (white arrow in **Figure S1h**) in in-vivo experiments. We assume that the unwrapped and B₀ drift-corrected phase-difference maps include both temperature information (which has a linear TE-dependence by a linear coefficient $\lambda$) and CSF-induced phase errors $\Delta\phi_{bias}$ (which is independent of TE):

$$\Delta\phi_{TEi\ unwrapped,\ drift\ corr} = \lambda \cdot TEi + \Delta\phi_{bias}.$$ [4]

Therefore, CSF-induced phase error $\Delta\phi_{bias}$ can be easily separated and removed by a linear least-square fit in each pixel $(x,y)$ based on Equation [4].

**Step 6. Multi-echo Combination**

In Step 6, a multi-echo combination (MEC) procedure is performed using a TNR-optimal weighted approach (i.e., weighted sum) according to Equation [5][38,39]:

$$\Delta T_{MEC} = \frac{\sum_1^n (S_{TEi}^2 \cdot TEi^2 \cdot \Delta T_{TEi})}{\sum_1^n (S_{TEi}^2 \cdot TEi^2)},$$ [5]

where $S_{TEi}$ is the signal magnitude at the $i$th TE, and $n$ is the number of TEs ($n$ = 4 in this study).

In brief, the entire six-step algorithm pipeline results in the multi-echo combined, signal loss and CSF artifacts corrected temperature imaging (i.e., MEC-Corr MRTI).

**2.3.2 Single-echo Thermometry for Comparison**

For comparison, single-echo data (e.g. TE1 or TE4) is also processed using the conventional single-echo PRF shift-based method. The phase unwrapping is conducted in a single-echo way[45]. That is, the complex-phase subtraction is first applied to create the phase difference map (identical to Step 1 in the multi-echo algorithm), then the temporal history of each pixel's phase is detected and unwrapped. If a discontinuity along the time dimension is encountered, then the phase is compensated for by adding $2\pi \cdot k$ $(k \in Z)$. The remaining procedures are similar to the proposed multi-echo-based algorithm, except that the single-echo algorithm does not include Step 4~6.

*2.4 MRgLITT Experiments*

**2.4.1 Phantom and Ex-vivo Heating Experiments**

Gel phantoms (N=2), ex-vivo pork samples (N=2), and ex-vivo swine brain samples (N=2) were utilized. These samples were placed in a plastic container, with four reference tubes filled with gel attached as the isolated references (**Figure 2a**). For the ex-vivo experiments, the container was filled with water for air isolation. Heating was applied using the laser applicator (MRI-Guided Laser Ablation System, Sinovation Medical, Beijing, China). The laser ablation probe used a 7-mm long cylindrical diffusing tip for heating, with saline coolant circulating through an outer 1.6-mm diameter catheter.

Two different heating modes were applied. One experiment used a continuous heating mode, with ablation lasting for 6, 3.5, and 2 minutes for the gel phantom, pork, and swine brain, respectively. The other experiment used an intermittent heating mode, with ablation lasting for about 60s and cooling for about 10s in one cycle. 12, 15, and 16 cycles were applied for the gel phantom, pork, and swine brain tissues, respectively. Gel phantoms were heated at 10 W, while pork and swine brain tissues were ablated at 8 W.

Data were acquired on a 3T MRI scanner (Ingenia CX, Philips Healthcare, Best, The Netherlands) with a 16-channel receiver coil using a multi-echo flyback GRE sequence with the following parameters: flip angle = 30°, TE = 5.5, 10, 14.5, 19 ms, TR = 22 ms, sampling matrix = 176 × 176, FOV = 200 × 200 mm$^2$, SENSE = 3, slice thickness = 5 mm, 3 slices parallel to the laser applicator, no gap, temporal resolution = 4 s / volume.

Two MR-compatible fiber-optic temperature probes (LUXTRON M920, Advanced Energy, USA) were inserted into the samples, with probe tips positioned approximately 8 mm away from the laser applicator (**Figure 2b**) in order to obtain accurate temperature measurements at these points as the ground truth. With the aid of the apertures on the top of the plastic mold (see **Figure 2a**, indicated by the white arrowhead), the two fiber-optic probes and the laser applicator were positioned within one imaging plane (in a transversal view). Because the fiber-optic probes were influenced by the laser applicator during heating due to band interference, only cooling stages were monitored by the thermometer. The MR-calculated temperature was obtained from a single voxel close to the fiber-optic probe. Then root-mean-square error (RMSE) was calculated to evaluate the temperature measurement precision. The temporal dynamics of the MR-calculated temperatures were compared with those of the thermometer-measured temperatures according to Equation [6]:

$$RMSE = \sqrt{\frac{1}{N}\sum_{t=1}^{N}(calculated_t - measured_t)^2} \ , \qquad [6]$$

where $t$ is the time point, $calculated_t$ is the MR-calculated temperature, $measured_t$ is the thermometer-measured temperature, and $N$ is the total number of time points at which the MR calculations and the thermometer measurements at the cooling stages are time-matched.

### 2.4.2 In-vivo Canine MRgLITT Ablation

Nine adult Beagle canines (N=9) underwent LITT ablation (**Figure 2c**). In-vivo ablations were approved by the local Institutional Animal Care and Use Committee (IACUC). Procedures regarding animal preparation and maintenance were detailed in **Supporting Information**. Before LITT ablation, 3D T1-weighted MR images (using the scanning parameters listed in **Table S1**) were first acquired to determine the location of the laser applicator. The canines were then heated at 8 W using the laser ablation system with different ablative doses, i.e., 50 s for Canine 01-05, 45 s for Canine 06, 60 s for Canine 07, 70 s for Canine 08, and 80 s for Canine 09. GRE temperature data were acquired on the same 3T MRI scanner with a 32-channel head coil using the multi-echo flyback GRE sequence with the following parameters: flip angle = 30°, TE = 4, 9, 14, 19 ms, TR = 22 ms, sampling matrix = 100 × 100, FOV = 130 × 130 mm$^2$, SENSE = 2.5, slice thickness = 4 mm, 3 slices parallel to the laser applicator, no gap, temporal resolution = 3 s / volume.

After LITT treatment, multi-contrast post-ablation MR images including T1-weighted Gadolinium-enhanced MRI, T2-weighted MRI, and T2-FLAIR were achieved using the scanning parameters listed in **Table S1** to determine the actual ablation range. The ablation volumes from the post-treatment MRI were then compared with those estimated from the temperature imaging using TE1 (4 ms), TE4 (19 ms), and MEC-Corr results based on the Arrhenius rate model[1,46].

Also, a qualitative visual assessment of temperature maps was performed by two experienced neurosurgeons using a 5-point scale ranking (see **Table S2**). Temperature images calculated from TE1 (4 ms), TE4 (19 ms), and MEC-Corr were blinded and assigned in a random order for evaluation. Three qualitative metrics were assessed: (1) Boundary smoothness of temperature hotspots, where rough boundaries indicate low TNR and smooth boundaries signify high TNR. (2) Temperature map artifacts, considering both signal loss-induced discontinuous points at the heating center and CSF-induced erroneous temperatures surrounding ventricles. (3) Overall thermometry reliability, taking into account the safety of the ablation operation as a whole, including a) whether rough boundaries would lead to misestimation of the ablation zone, b) whether incorrect temperature measurements at the heating center could result in LITT overheating or laser tip vaporization or carbonization, and c) whether CSF-induced temperature artifacts could mislead surgeons.

The reviewers' scores from TE1, TE4, and MEC-Corr were compared using Wilcoxon signed-rank test. The agreement between the two reviewers was assessed with intra-class correlation coefficients (ICCs). All statistical analyses were performed using SPSS software (version 21.0, SPSS Inc., Chicago, IL, USA). P <0.05 was considered statistically significant.

# Results

### *Evaluation of Temperature Imaging Signal Voids*

**Figure 3** reveals that LITT heating can cause a rapid image magnitude decrease at the center of the ablation region in the in-vivo canine brain, which can be an important cause of the MRTI signal void. The image signal loss recovers gradually after the heating is off, indicating that the signal loss is mainly related to the high temperature. Specifically, when the laser is on, the signal magnitude within the central heating region (i.e., Pixel 1, blue line) drops close to the noise level (Pixel 4, violet line). Correspondingly, the phase value of Pixel 1 exhibits irregular changes, differing from that of Pixel 2 (orange line). Once the laser is off, the phase signal of Pixel 1 gradually returns to normal.

**Figure 4** illustrates an inverse relationship in the in-vivo canine brain between the temperature change (orange line) and the T2* value (blue line) measured by the multi-echo GRE sequence, whether the target pixel is near  (**Figure 4a**) or far from (**Figure 4b**) the laser tip. In **Figure 4a**, the measured T2* value drops quickly once the laser is on, then approaches a plateau, and finally recovers after the heating is off. It matches the temperature change, which shows a quick increase at the beginning of the laser heating. Also, as the T2* value is decreased to < 10 ms at Time Point # 108 s in (a), the GRE image on the left (TE4 = 19 ms) exhibits severe signal loss at the heating center. In **Figure 4b**, the T2* curve changes more subtly as the temperature change is minor. Note that the initial T2* value near the laser tip (**Figure 4a**) is lower than the value approximately 8 mm away from the laser tip (**Figure 4b**) before heating, which is likely due to the $B_0$ inhomogeneity caused by the susceptibility difference between the laser tip and tissue interface.

### *Temperature Imaging Using Different TE Values*

**Figure S2** and **Movie S1** demonstrate the results of the single-echo thermometry at various TE settings. Compared to images at longer TEs (e.g., TE4), temperature imaging at shorter TEs (e.g., TE1) exhibits rougher boundaries around the hotspots in both ex-vivo and in-vivo studies, indicating lower TNR and less precise temperature measurement. On the other hand, images at shorter TEs (e.g., TE1) have less T2* decay and thus can be used to calculate temperature maps at the heating center thanks to the slighter signal loss on the magnitude images. The phase wraps also decrease as TE gets shorter, suggesting that phase unwrapping is either unnecessary or simpler. Additionally, in either TE setting, the in-vivo temperature maps demonstrate high temperatures within the third ventricle (indicated by the yellow arrow in **Figure S2**). In particular, the high temperatures are more apparent when using shorter TE values. Since high temperatures appear before laser heating (see **Movie S1**), they are more likely caused by CSF-induced phase errors rather than by thermal changes occurring when the surrounding brain parenchyma is heated.

### *Performance of the Multi-echo Thermometry*

Figure 5 demonstrates the performance of the proposed multi-echo-based temperature data processing algorithm on Canine 01 with 3 example pixels. **Figure 5a** shows the output of Steps 1-3 from Pixel 1, a normal voxel within the heating zone that is unaffected by the signal loss or CSF pulsation but needs phase unwrapping due to the large temperature increase. As indicated by the black arrow in (i), phase wrapping occurs at TE3 and TE4 as the temperatures significantly rise. The unwrapped phase difference at the longer TEs (i.e., TE2 - TE4) can be estimated based on the first echo according to Equation [2], as illustrated by the dashed lines in (ii). The third subfigure (iii) demonstrates the corresponding multi-echo-based phase unwrapping results of Step 2. The efficacy of $B_0$ phase drift correction (Step 3) is illustrated in (iv).

**Figure 5b** shows the result of Step 4 (i.e., Correction of Signal Loss-Induced Error). Pixel 2 is affected by the MRTI signal void and shows obvious temperature errors (black arrow in v) at longer TEs (i.e., TE2 - TE4). The temperature errors are well fixed by the correction in Step 4 based on the TE1 signals (shown in vi).

**Figure 5c** illustrates the results of Step 5 (i.e., Correction of CSF-Induced Error). As shown, Pixel 3 is influenced by CSF pulsation. Subfigures (vii and ix) demonstrate the phase difference changes and the corresponding temperature changes without CSF error correction, while subfigures (viii and x) display the results with CSF error correction. The four TEs have similar CSF-induced phase fluctuations, as indicated in (vii), and cause shorter TEs to experience larger temperature variances (see ix, TE1, blue line). The CSF phase bias is eliminated by the linear least-square fit in Step 5, as shown by the corrected phase (viii) and temperature (x) plots.

### *Phantom and Ex-vivo Experiments*

Zoomed-in temperature maps from the intermittent heating experiment on the ex-vivo pork sample are shown in **Figure 6**. Six example images are selected from the 300 frames (4 s / frame) acquired during the ablation. The first and second rows are the temperature maps calculated from the single-echo data (i.e., TE4 = 19 ms) using the conventional single-echo-based phase unwrapping and the proposed multi-echo-based phase unwrapping algorithm, respectively. Due to the signal loss, the traditional single-echo-based phase unwrapping technique corrupts the pixels at the heating center, and the errors (red arrowhead) cannot be recovered even when the laser is off (Time # 750 s). This can be explained by the fact that the conventional single-echo-based phase unwrapping method is applied along the time dimension for phase wrap detection. If the phase-difference map at the current frame is wrongly unwrapped, all the following frames are influenced. The proposed multi-echo-based phase unwrapping algorithm, on the other hand, operates along the multi-echo dimension, therefore avoiding the phase errors induced by the previous frames. The third row is the signal loss-corrected temperature maps at TE4, with the corrupted pixels at the heating center recovered. The last row is the temperature maps via MEC-Corr processing. Note that the final MRTI shows more uniform temperatures at the hotspot thanks to the multi-echo combination. **Movie S2** (a-c) displays TE1, TE4, and MEC-Corr temperature maps from the gel phantom, pork, and swine brain samples, respectively. The temperature maps also demonstrate that MEC-Corr performs better than single TE1 and single TE4, as TE1 results show rough hotspot boundaries (indicating inadequate TNR) and TE4 results suffer from MRTI signal voids (providing insufficient temperature accuracy).

**Table 1** lists the calculated RMSE values from single TE1, single TE4, and MEC-Corr in experiments with the two different heating modes. **Figure S3** illustrates the example heating curves for the gel phantom (in the intermittent heating mode) and the pork tissue (in the continuous heating mode) based on the MR calculated (i.e., MEC-Corr, black lines) and optic-measured temperatures (i.e., red lines), respectively. As demonstrated in **Table 1**, the temperature values near the fiber-optic probes are not corrupted by the MRTI signal voids in the gel phantom experiments, thus yielding an averaged RMSE value of 0.478 ℃ for TE1, 0.22 ℃ for TE4, and 0.175 ℃ for MEC-Corr, respectively. In the pork experiments, the averaged RMSE values for TE1 and MEC-Corr are 0.775 ℃ and 0.403 ℃, whereas the RMSE values for TE4 near the right (R) probe reach 4.81 ℃ in the continuous heating experiment and 30.6 ℃ in the intermittent heating experiment, indicating that TE4 temperature measurements are contaminated by signal voids. The RMSE value of 30.6℃ in the intermittent heating experiment also suggests that the voxel undergoes incorrect phase unwrapping (as the phase change of $2\pi$ refers to a temperature change of around 41 ℃ for TE4 = 19 ms). In the brain tissue experiments, the average RMSE values for TE1 and MEC-Corr are 0.675 ℃ and 0.443 ℃, whereas the RMSE value for TE4 near the left probe reaches 39.3 ℃ in the continuous heating experiment, which also suggests the signal loss-induced incorrect phase unwrapping.

### *In-vivo MRgLITT Ablations*

**Figure 7** shows example temperature maps using single TE1, single TE4, and MEC-Corr from Canine 03. The temperature images are selected from the 100 frames (3 s / frame) acquired during laser heating and are overlaid onto the post-ablation T2-weighted images. In this experiment, the heating zone is located close to the ventricles. The first row (TE1 = 4 ms) shows artifactually high temperatures within the third ventricle (red arrowhead), indicating that short TE acquisitions are sensitive to CSF motion. The second row (TE4 = 19 ms) demonstrates that TE4 provides smoother heating boundaries (i.e., greater TNR) than TE1, but pixels within the heating center are corrupted due to MRTI signal voids. Additionally, the CSF-induced artifacts within the third ventricle (red arrowhead) are still visible on TE4 temperature maps. In contrast, MEC-Corr results provide temperature maps free of the artifacts caused by either CSF or signal loss. **Movie S3** (a-c) displays the temperature maps of Canine 01–03 throughout the heating process from TE1, TE4, and MEC-Corr. This supports the assertion that MEC-Corr results outperform single TE1 and single TE4 results, providing higher temperature measurement accuracy.

**Figure 8** illustrates the thermal ablation results of in-vivo MRgLITT treatment on three example canines (i.e., Canines 01-03) that were heated at 8W for 50 seconds. The zoomed-in subfigures in **Figure 8** demonstrate satisfactory consistency between the ablation volumes estimated from MRTI (i.e., MEC-Corr) and the ablation volumes measured from the post-ablation MRI (i.e., T2w, FLAIR, and MPRAGE). **Figure S4** shows the post-ablation T2w images (first row) and the MEC-Corr MRTI predicted ablation volumes (second row, shown in red) from the other six canines (i.e., Canines 04-09) with different ablative laser doses (8W, tens of seconds) at different locations. The laser ablation areas range from less than 40 mm$^2$ to nearly 120 mm$^2$, for different locations and different laser applied times. The images show that the ablation estimates from MRTI (i.e., MEC-Corr) match well with the post-ablation MR assessments. Furthermore, **Figure S5** demonstrates the thermal ablation results estimated by either single TE1, single TE4, or MEC-Corr MRTI from two example canines, i.e., Canine 02 (unaffected by the MRTI signal voids) and Canine 05 (affected by the MRTI signal voids). Notice how the accuracy of the TE1 ablation estimation in **Figure S5a** and the TE4 ablation estimation in **Figure S5b** is impacted by temperature errors caused by signal loss and CSF flow, respectively. In addition, the rougher hotspot boundaries in TE1 result in less precise ablation margins in both **Figure S5a** and **S5b**. Precise temperature imaging, however, enables MEC-Corr to produce accurate ablation estimates in both canine experiments.

Figure 9 illustrates the scores rated based on the calculated temperature maps using single TE1, single TE4, and MEC-Corr from the two neurosurgeons. The ICCs value of 0.746 indicates moderate agreement between the two reviewers. The MEC-Corr results demonstrate the best scores in all three metrics: (1) TE1 vs. MEC-Corr MRTI: smoothness 2.17 vs. 4.11 (p=0.0039), artifacts 2.83 vs. 3.89 (p=0.0313), reliability 3.22 vs. 4.00 (p=0.0078). (2) TE4 vs. MEC-Corr MRTI: smoothness 3.06 vs. 4.11 (p=0.0156), artifacts 2.67 vs. 3.89 (p=0.0078), reliability 3.17 vs. 4.00 (p=0.0313). For boundary smoothness, MEC-Corr (4.11) > TE4 (3.06) > TE1 (2.17) indicates that TNR improves with TE, and the TNR-optimal weighted echo combination obtains the best TNR. For temperature artifacts, both TE1 and TE4 show poor performance (scores < 3), majorly due to the CSF flow artifacts (which primarily affect TE1) and the signal loss (which primarily influence TE4). On contrary, MEC-Corr results receive the best rating (3.89) with the corrections from Steps 4 and 5. Additionally, compared to TE1 and TE4, MEC-Corr results show the best clinical reliability.

## Discussion

Accurate temperature monitoring is crucial for MRgLITT. However, signal loss can introduce temperature artifacts in PRF shift-based thermometry. To address this issue, in this work, we adopt the multi-echo GRE acquisition strategy given that GRE images at shorter TEs are less affected by signal artifacts. The multi-echo GRE sequence offers complementary information from the varied TEs without increasing the scan time because only shorter TEs are added. The T2* variations depending on the T2* measurement can be covered by the TE selection range for the multi-echo acquisition. Furthermore, we also develop a multi-echo-based correction algorithm to minimize the signal loss-induced thermometry errors, which also includes a new phase unwrapping method and a CSF pulsation correction method. These further improve the accuracy of the temperature mapping method. The results demonstrate that the multi-echo-based method can correct the MRTI signal voids and achieve higher accuracy when compared with the traditional single-echo method in either phantom, ex-vivo, or in-vivo experiments.

The novelty of this study includes the investigation of the cause of the MRTI signal void and the development of the multi-echo-based algorithm to address the signal void in MRgLITT. In this study, we discovered that the T2* values can be greatly shortened at high temperatures during laser heating (see **Figure 4**), which is a significant contribution to the MRTI signal void. As a result, a combination of the varied TE values (i.e., short TE for low T2* with high temperature, and long TE for high T2* with low temperature) is utilized. Although the individual steps of the proposed multi-echo-based algorithm have been covered in previous research, the pipeline itself is innovative and effectively eliminates signal loss-induced artifacts and CSF-induced temperature errors, while enhancing the robustness of phase unwrapping in the presence of artifacts.

The GRE signal loss that appears in MRgLITT can be attributed to several factors. First, the experiment in this study reveals that T2* within the hotspot decreases greatly during laser heating. This inverse relationship of T2* with temperature is also mentioned in another study[40]. The large temperature change across the voxel, i.e., the through-plane and in-plane temperature gradient can be one of the main factors that contribute to the reduction of T2*. Another factor for the T2* decrease might be the formation of gas bubbles during ablation[47-51]. Furthermore, T2* shortening may be related to tissue thermal swelling[52], which can cause a complicated, localized, three-dimensional warping of the phase distribution. Local susceptibility changes can also be a contributing factor, which is influenced by the type of tissue[53] and the pattern of the delivered temperature rise[54]. All factors can lead to a T2* decrease, which eventually results in a severe signal void in MRTI. Second, T1 increase caused by high temperatures[55] might also contribute to signal loss. According to earlier research, a temperature rise during LITT can increase T1, thus resulting in a distinct signal drop within the heated area[56,57]. In future work, the causes of signal loss during LITT treatment need to be further investigated.

The set of TE values employed in the multi-echo acquisition in this study works well based on the T2* measurements acquired during MRgLITT treatment. In PRF temperature imaging, TE = T2* offers the best TNR and the greatest temperature measurement precision. Therefore, short TE (i.e., TE = 4~10 ms) is applied for short T2* (i.e., T2* ≤ 10 ms) voxels at high temperatures (**see Figure 4**) to achieve accurate temperature measurements that are free from T2* decay-induced signal loss, while long TE (i.e., TE = ~20 ms) is applied for long T2* (i.e., T2* = 25~40 ms) voxels at low temperatures (**see Figure 4**) to strike a balance between high TNR and high acquisition efficiency (i.e., short acquisition time).

The proposed multi-echo thermometry integrates the advantages of both short TE and long

TE. The results show that short TE (~ 4ms) is susceptible to CSF flow artifacts (e.g., surrounding the ventricles in in-vivo canines) while long TE (~ 19 ms) is more likely to be influenced by signal loss-induced temperature artifacts. However, the proposed multi-echo-based technique can reliably reduce both types of artifacts and increase the accuracy of temperature measurements. In the in-vivo canine ablation experiments, the multi-echo thermometry shows smoother hotspot borders and fewer MRTI artifacts, and improves the consistency between the ablation zone estimated from intraoperative MRTI and that measured from post-ablation MRI (see **Figure S5**). Overall, the proposed multi-echo thermometry provides more accurate temperature imaging, which can benefit the clinical practice of MRgLITT from several aspects. Firstly, better prevention of tissue vaporization or carbonization surrounding the laser fiber tip is made possible by accurate temperature monitoring at the heating center. Secondly, the multi-echo-based thermometry also boosts the consistency between the ablation region estimated by MRTI and that measured from the post-ablation MRI in cases when significant MRTI signal voids happen. In traditional single-echo thermometry, the ablation estimation can be corrupted by these signal voids (e.g., in **Figure S5**). Despite the strengths, the proposed data processing pipeline can be further improved. For instance, to simplify algorithm implementation, here the temperatures at TE1 are directly used to replace those at longer TEs in cases where voxels are corrupted. In future work, the correction of signal loss-induced artifacts can be merged in Step 6 by setting the weights of the corrupted voxels to zero, so that the temperatures of longer TEs (e.g., TE2) can also be utilized. Thus, even at the voxels with signal loss, a higher TNR can still be obtained.

Apart from combining the intrinsic advantages of both short and long TEs, the multi-echo acquisition also enables a new phase unwrapping algorithm for huge temperature changes. The conventional phase unwrapping procedure along the temporal dimension can be problematic when signal loss appears on GRE images. The proposed multi-echo-based phase unwrapping is resistant to the influence of signal artifacts, as the phase unwrapping is done along the echo dimension rather than the temporal dimension, and the first TE (i.e., TE1) applied for unwrapped phase estimates is typically unaffected by the signal loss. Phase unwrapping using multi-echo GRE was also implemented by Svedin et al[40]. The method calculates the phase difference between adjacent echoes for phase unwrapping under the assumption that the phase increment between echoes is smaller than $2\pi$. Our proposed unwrapping method is based on the phase and TE relation, and thus no additional conditions between echoes are required.

Besides the modified phase unwrapping, our proposed multi-echo-based algorithm also integrates effective correction of CSF pulsation artifacts. CSF-induced temperature artifacts are more noticeable on shorter TEs (see **Figure S2b**). This can be explained by the fact that the CSF-induced phase error is TE-independent[58]. For instance, if the CSF motion induces a constant phase error $\Delta\phi_{CSF}$ (~0.1 rad), it would cause a temperature change of ~0.65 ℃ according to $\Delta T_{CSF} = \Delta\phi_{CSF}/(\alpha\gamma B_0 TE)$ when TE = 19 ms. But the error reaches a higher value of 3.11 ℃ (by a factor of 4.75) when TE is shortened to 4 ms. In this study, we employ the multi-echo-based approach to correct the CSF-induced temperature errors, which is inspired by a previous work that uses multi-echo (3TEs) data to differentiate BOLD and non-BOLD signals in fMRI[58]. Based on the concept of TE-dependence, non-BOLD-like components[58] (i.e., TE-independent components) related to motion, pulsatility, and other unwanted factors are effectively removed. Besides CSF flow correction, the multi-echo-based motion correction method may also help correct the phase errors caused by pulsatile blood flow, and thus cardiac triggering[59] or flow compensation encoding is no longer required for accurate temperature imaging. One drawback of the correction of CSF-induced artifacts is that the voxel-based least-square fitting is time-consuming. On a computer with a 3.4GHz Intel Core i7 CPU and 32GB of RAM, it takes around 0.25s to run the fitting in MATLAB for one frame and one slice. This can be skipped if the heating zone is free of CSF flow, or can be replaced by other motion correction methods like the multi-baseline method[21]. Still, the effectiveness of the multi-echo-based motion correction should be further investigated.

Over the years, multiple applications of multi-echo thermometry have been proposed. One of the primary objectives is to improve the TNR, particularly when the readout bandwidth is increased to reduce the geometric distortions and blurring induced by off-resonance[39,40]. These works typically use around ten echoes and therefore the readout bandwidth can be increased several times. However, to compensate for the signal loss-induced phase errors correctly, the readout bandwidth in this study is comparable with that in the conventional single-echo acquisition to achieve as high TNR as possible at every single TE. Additionally, multi-echo thermometry has been utilized to recover the temperature imaging near metallic ablation probes using the dual-echo z-shimming method[42]. However, this method[42] cannot be employed directly for MRgLITT, because the heat-induced signal loss is changing rather than static (see **Figure 3**), which necessitates dynamic z-shimming but requires complex implementation. Instead, the multi-echo acquisition in this study (with TE1 ~4 ms) offers an alternative solution.

Besides using multi-echo thermometry, other approaches can also help alleviate or avoid signal loss-induced temperature artifacts. One strategy is to reduce the slice thickness, thus decreasing the through-plane gradient-induced $T2^*$ dephasing but at the cost of SNR. Another strategy is to choose TE carefully so that the MRTI signal voids can be prevented and sufficient TNR can be kept. However, a single TE usually fails to satisfy all measured voxels throughout the heating process, especially when different laser heating powers are applied for one treatment. Additionally, the specific anatomical structures in the human brain (such as sinuses and ear canals) and the use of auxiliary equipment (such as the fixation apparatus) may also introduce additional $B_0$ inhomogeneity, thus lowering initial $T2^*$ before laser heating and complicating temperature measurements. Therefore, multi-echo thermometry is helpful in clinical MRgLITT applications.

The multi-echo GRE sequence is commonly used in clinical scans, making it simple to implement the proposed multi-echo thermometry on various MR platforms. Preliminary MRgLITT results using the proposed method have been demonstrated in epilepsy patients, which show the superiority of multi-echo thermometry over conventional single-echo thermometry[60]. In the future, systematic clinical studies are warranted to clarify the accuracy of MRgLITT using the proposed multi-echo thermometry.

## Conclusions

In conclusion, T2* shortening can be one important cause of the MRTI signal voids and thus demand for a multi-echo acquisition covering both short and long echo times. A multi-echo-based temperature mapping algorithm with suitable TE assignments is therefore presented to correct the MRTI signal voids. A more reliable phase unwrapping approach and a new CSF motion correction method are two additional advantages of the multi-echo-based approach. The overall multi-echo thermometry is validated via a series of ex-vivo and in-vivo experiments. The results show that compared with single-echo thermometry, the multi-echo-based approach can obtain fewer temperature artifacts and smoother hotspot boundaries. With its potential for safer LITT treatment and compatibility with commercial MR scanners, multi-echo thermometry holds promise for clinical application.

## Acknowledgments

This work was supported by the National Key R&D Program of China (2022YFC2405303) and the Beijing Municipal Natural Science Foundation (L192006).

# Figures

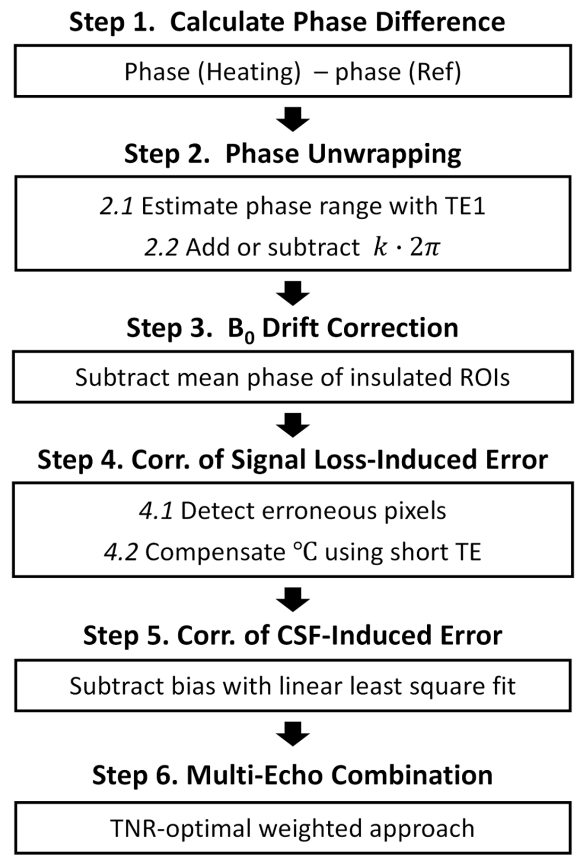

**Figure 1.** Flow diagram of the proposed multi-echo-based data processing pipeline. *Abbreviations*: Corr., correction; Ref, reference.

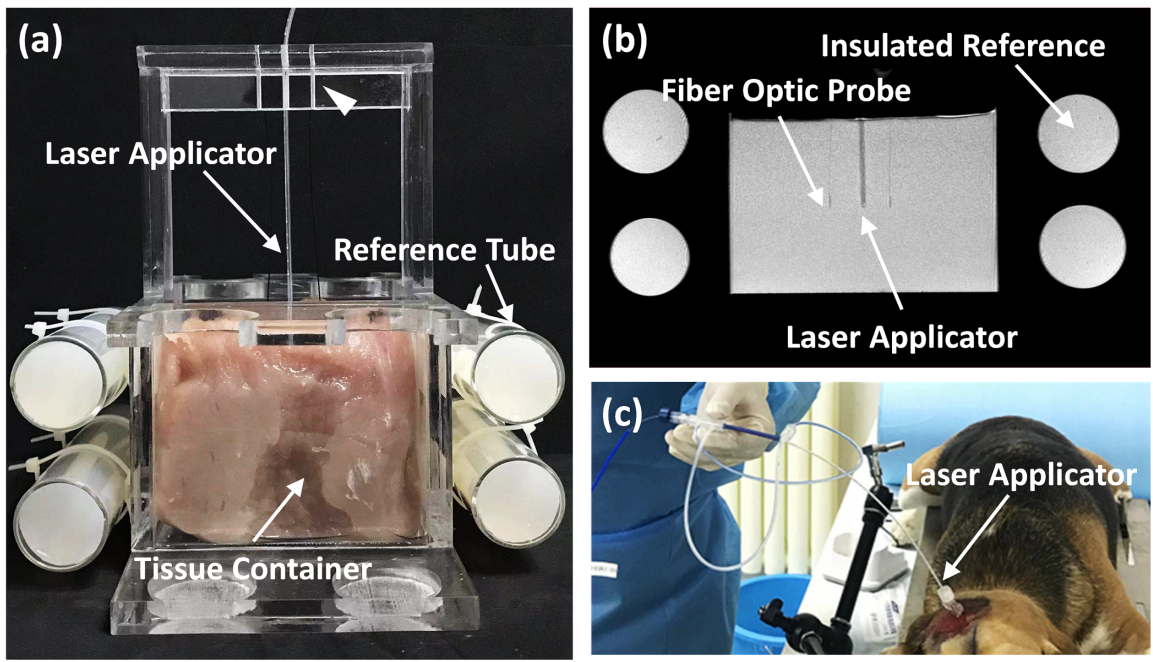

**Figure 2.** Experiment setup for the phantom, ex-vivo, and in-vivo experiments. (a) Pork tissue was placed in a plastic container with four reference tubes fixed around it. The apertures on the top of the plastic mold (white arrowhead) were utilized to facilitate the placement of the laser applicator and fiber-optic probes. (b) T2-weighted images showing the positions of the laser applicator and two fiber-optic probes in a gel phantom. (c) Picture showing the placement of the laser applicator on a Beagle skull. The laser applicator was positioned through a plastic cranial bolt.

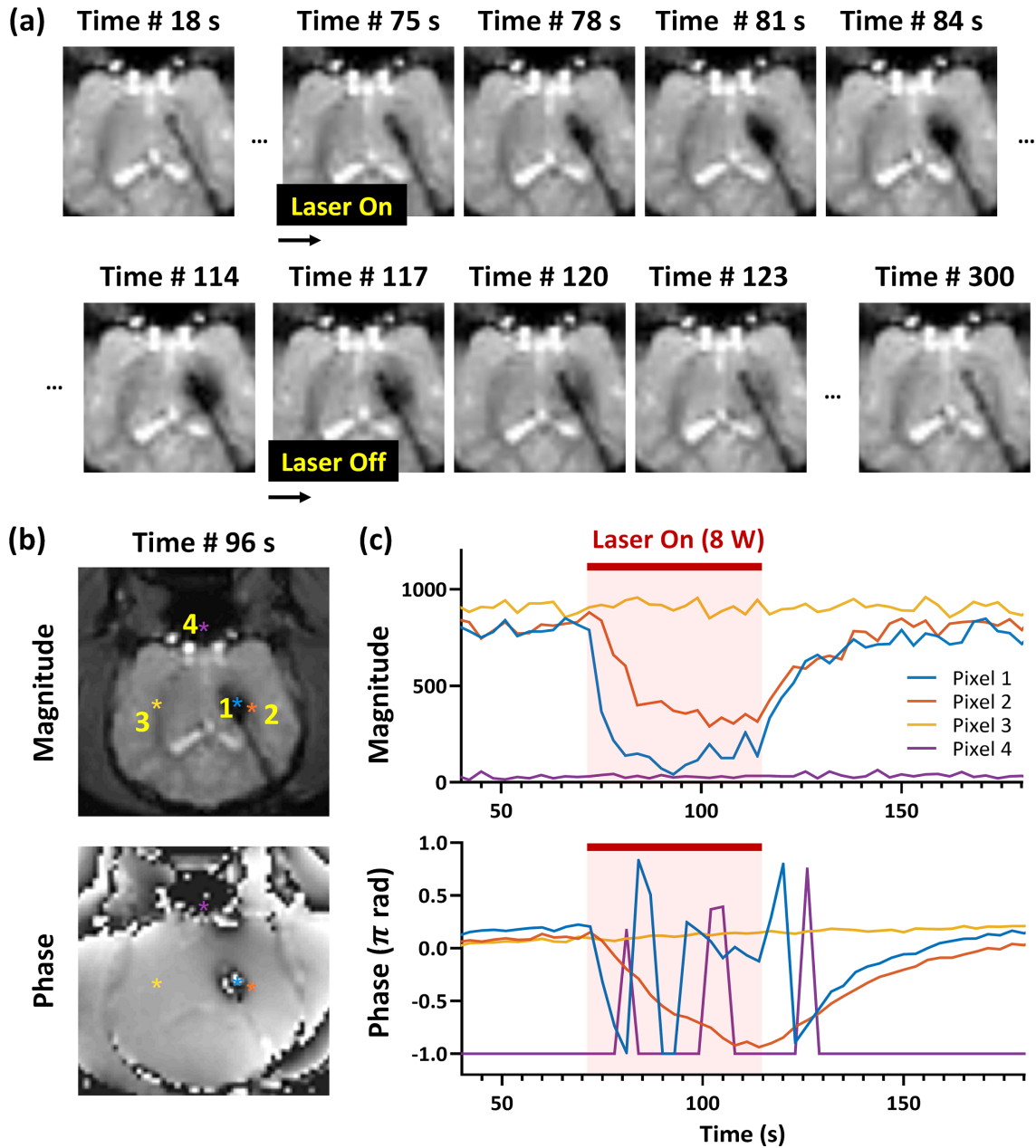

**Figure 3.** Magnitude signal loss during laser heating on the GRE images (TE = 19 ms) of an example canine brain (i.e., Canine 01). The zoomed-in MR images in (a) show signal loss once the laser is on, and the signal loss recovers quickly after the laser is off. Four pixels (1-4) marked as stars in (b) are selected for illustration. The corresponding magnitude and phase change of pixels 1-4 are then plotted in (c) as a function of time. Pixel 1: a voxel located in the signal void within the heating zone; Pixel 2: a voxel unaffected by signal loss within the heating zone; Pixel 3: a voxel not heated; Pixel 4: background noise.

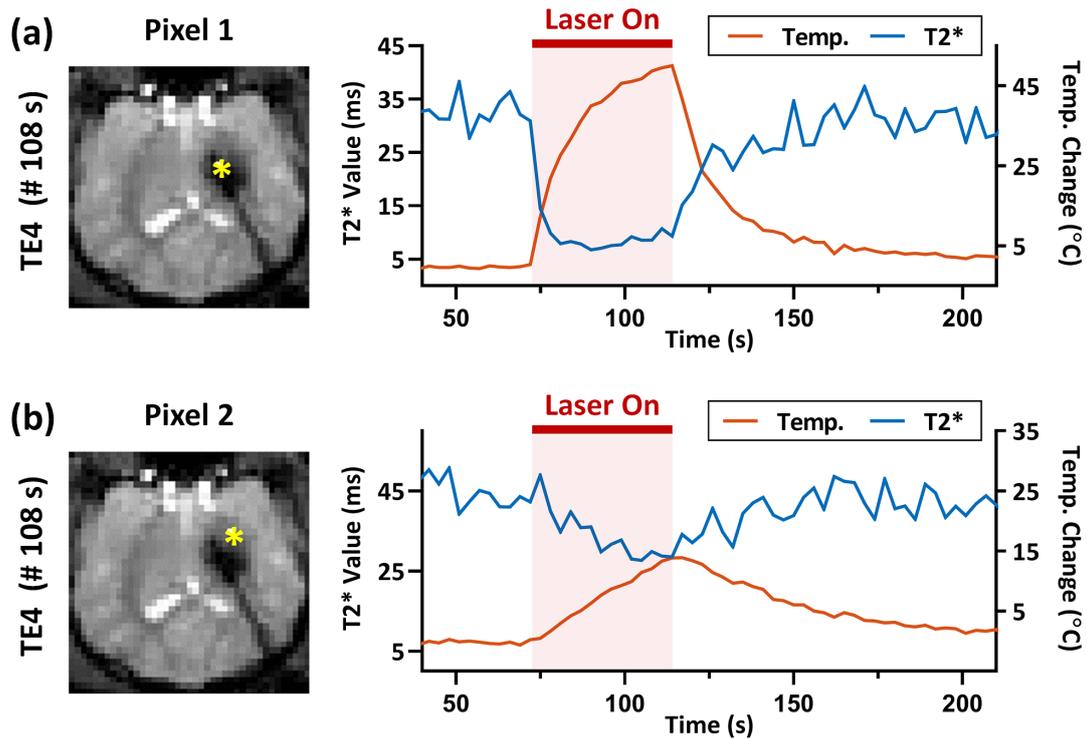

**Figure 4.** *Left*: The locations of the two measured pixels (indicated by the yellow stars), demonstrated on the GRE images of Canine 01 (TE4 = 19 ms, Time Point = 108 s). *Right*: Measured T2* change (blue line) and temperature change (orange line) obtained via MEC-Corr during LITT. (a) One pixel near the laser tip with a large temperature change (the temperature is not corrupted by the MRTI signal void); (b) one pixel approximately 8 mm away from the laser tip with a minor temperature change.

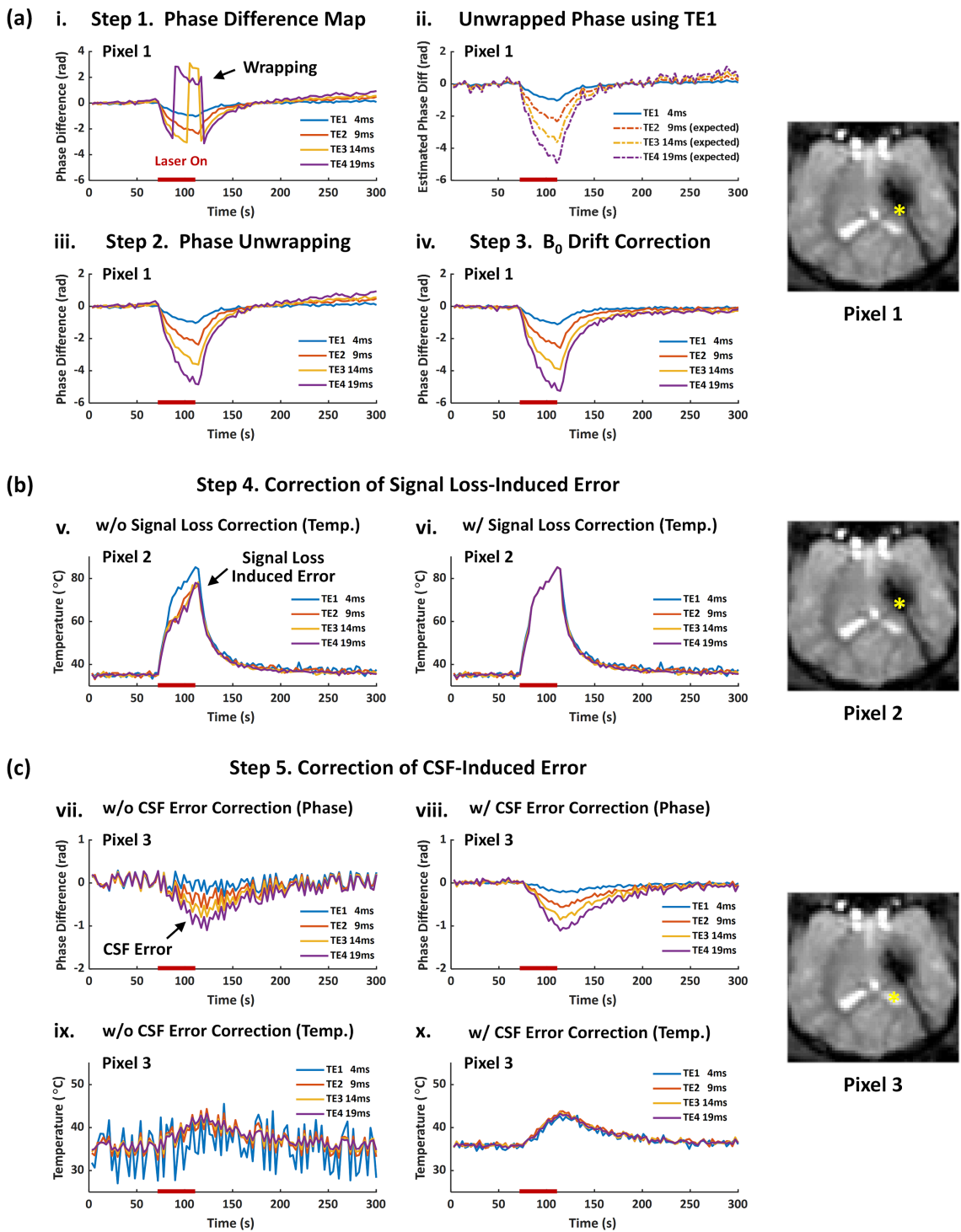

**Figure 5.** Performance of the proposed multi-echo-based algorithm at three example pixels. (a) illustrates how the phase unwrapped, $B_0$ drift-corrected phase difference data is obtained based on Steps 1-3, with an example pixel near the heating center (not corrupted by the signal loss or CSF artifact). The first subfigure (i) demonstrates the residual phase wrapping (black arrow) that occurs at high temperatures. The second subfigure (ii) demonstrates the expected unwrapped phase range of longer TEs (dotted line) based on TE1, which gives a clear indicator of the phase

unwrapping for the large temperature changes in (i). (b) illustrates how the signal loss-induced error is corrected, with an example pixel within the heating zone that is corrupted by the MRTI signal void. Temperature errors at longer TEs (black arrow) are corrected using the shortest TE1. (c) illustrates how CSF-induced error is corrected. The plots in the first and bottom rows are the phase difference (vii, viii) and corresponding temperature changes (ix, x) as a function of time. The 4 TEs have similar values of the CSF-induced phase fluctuations (black arrow), which causes more pronounced temperature errors at shorter TEs (i.e. TE1, blue line) without CSF error correction (ix). The CSF error-corrected results (x) show smoother temperature curves.

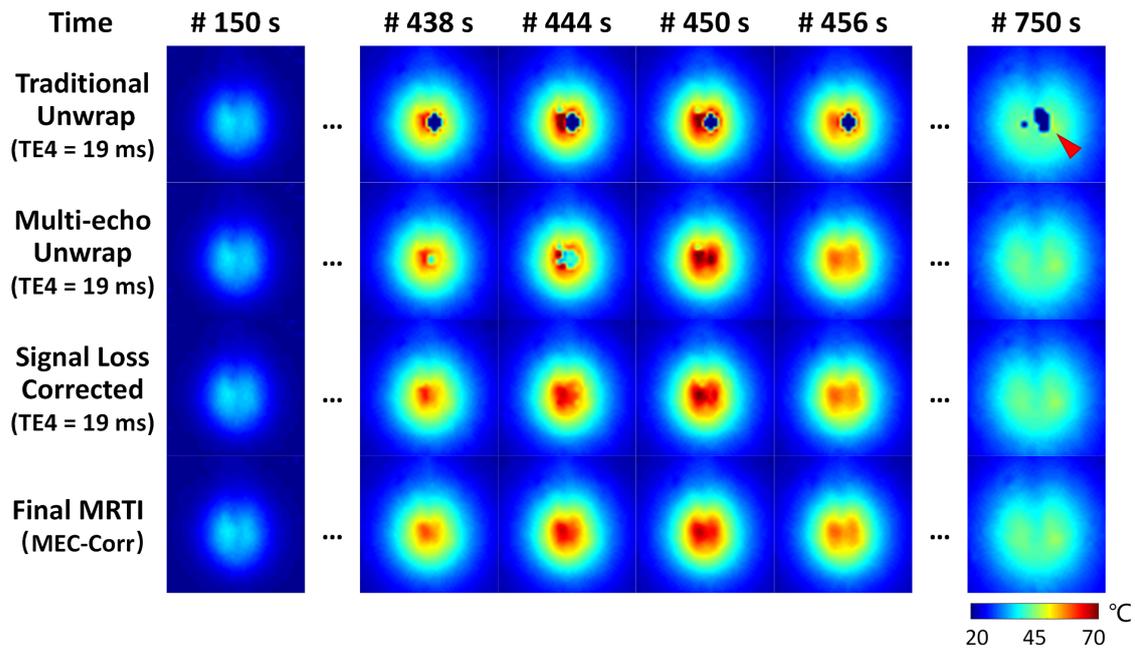

**Figure 6.** Example temperature images of the ex-vivo pork tissue during LITT heating. From top to bottom: zoomed-in temperature maps calculated from 1) the conventional phase unwrapping along the time-dimension (demonstrated on data using TE4 = 19 ms); 2) the proposed multi-echo-based phase unwrapping; 3) the proposed multi-echo-based phase unwrapping and the correction of the signal loss-induced error; 4) the whole multi-echo algorithm (i.e., MEC-Corr results). Note that TE4 temperature maps calculated using the conventional phase unwrapping method demonstrate noticeable unwrapping errors at the hotspot that persist after the heating is off (Time # 750 s, red arrowhead). However, the temperature maps that employ the proposed multi-echo-based phase unwrapping can revert to accurate measurements after LITT heating is finished. The final MEC-Corr images demonstrate improved uniformity at the hotspot compared to TE4 results (the third row).

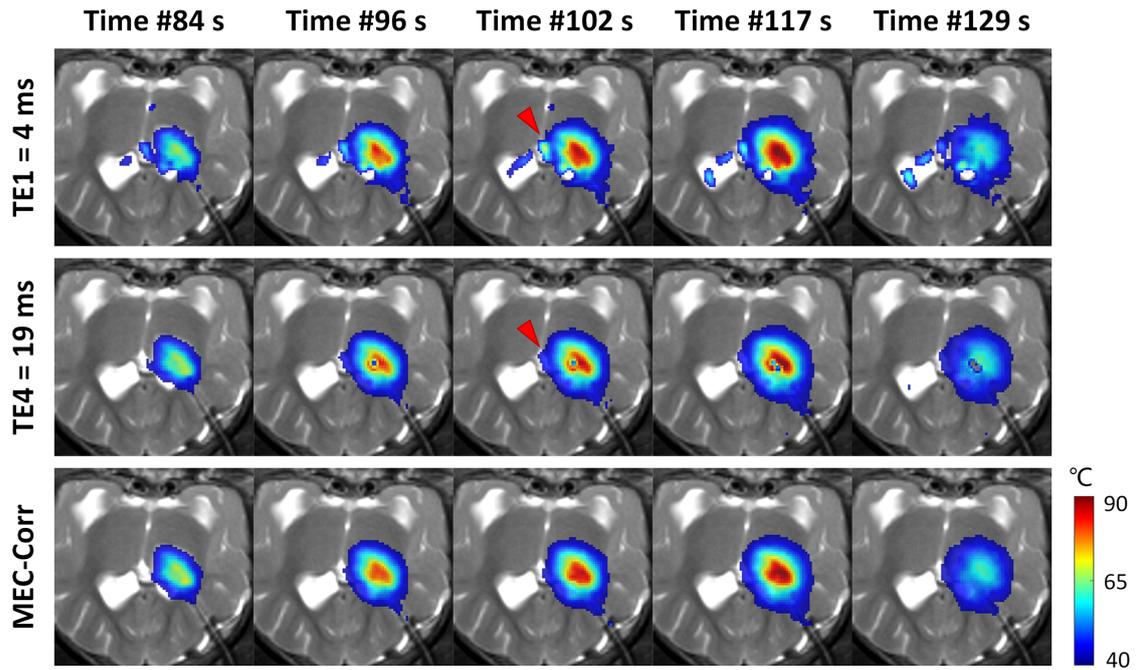

**Figure 7.** Example temperature images of Canine 03 during LITT heating. From top to bottom: 1) calculated from TE1 data via the traditional single-echo PRF algorithm; 2) calculated from TE4 data via the traditional single-echo PRF algorithm; 3) calculated from the 4-echo data using the proposed multi-echo thermometry. MEC-Corr MRTI provides effective correction of the signal loss-induced artifacts at the heating center and suppression of the CSF flow artifacts near the ventricles (indicated by the red arrowheads).

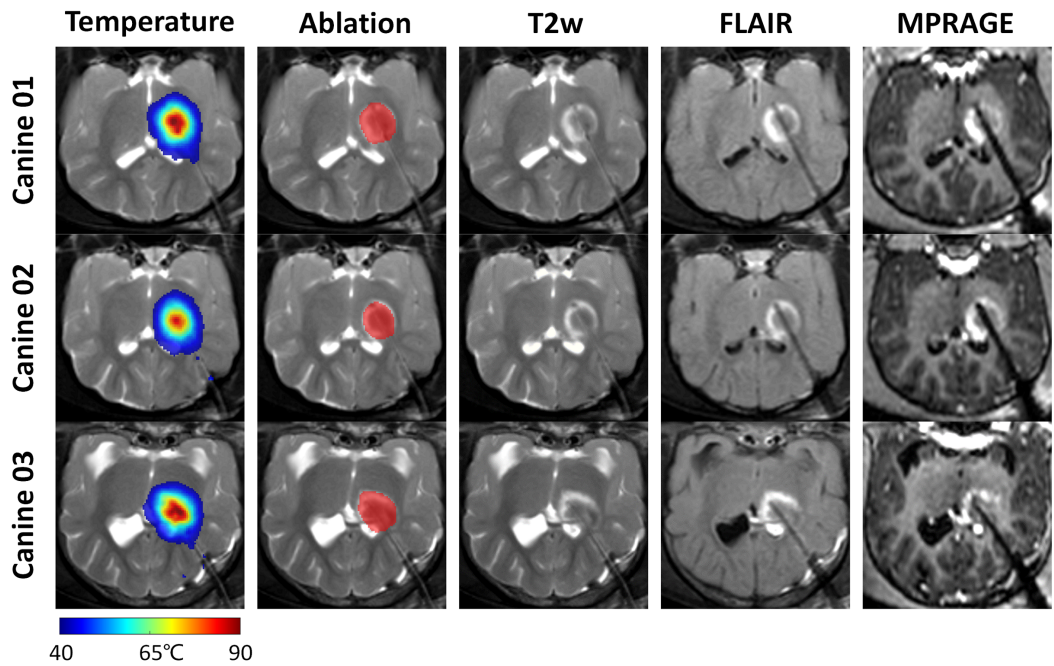

**Figure 8.** Thermal ablations of 3 example canines (i.e., Canine 01-03). From left to right: 1) MEC-Corr temperature images at the most intense heating; 2) estimates of thermal ablations shown in red from the MEC-Corr temperature measurements; 3) post-ablation T2-weighted MR images; 4) post-ablation FLAIR images; 5) post-ablation gadolinium-enhanced MPRAGE (i.e., T1-weighted) images.

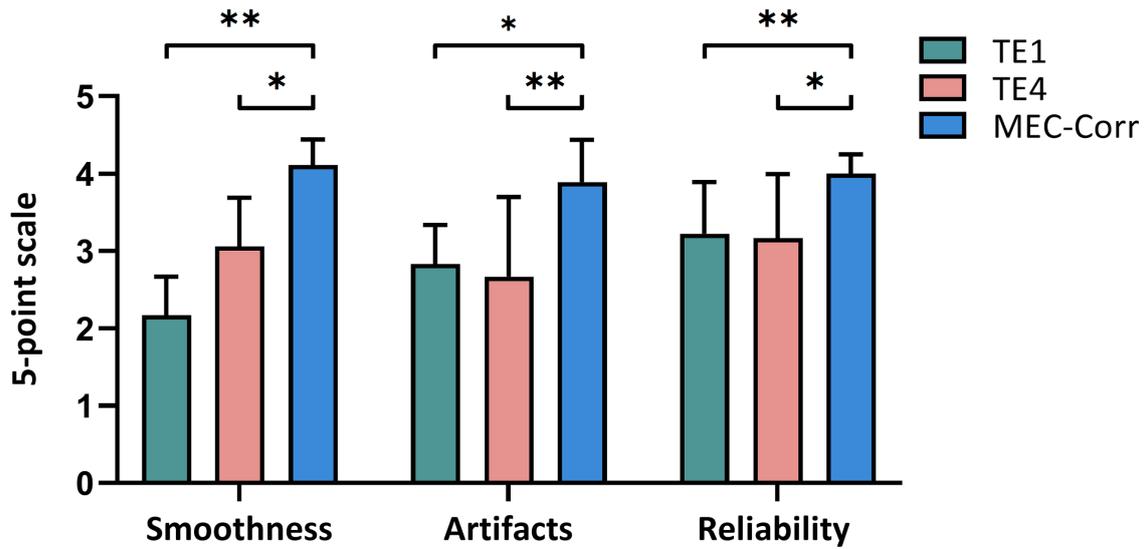

**Figure 9.** Image-quality assessment of the temperature boundary smoothness, temperature map artifacts, and overall measurement reliability for single-echo (TE1 and TE4) and multi-echo (MEC-Corr) thermometry. The Wilcoxon signed-rank test results are marked above each pair of the groups. *Abbreviations*: TE1, TE = 4 ms (shortest TE); TE4, TE = 19 ms (longest TE); MEC-Corr, multi-echo combined and artifacts corrected MRTI.

# Tables

**Table 1.** Comparison of the MR calculated and the fiber-optic-measured temperatures for TE1 (4 ms), TE4 (19 ms), and MEC-Corr MRTI, respectively.

| RMSE (℃) | Gel Phantom | | Pork | | Swine Brain | |
|---|---|---|---|---|---|---|
| H. Mode | Cont. H. | Interm. H. | Cont. H. | Interm. H. | Cont. H. | Interm. H. |
| TE1 | 0.36 (R) | 0.48 (R) | 0.78 (R) | 0.61 (R) | 0.57 (R) | 0.80 (R) |
| | 0.41 (L) | 0.66 (L) | 0.88 (L) | 0.83 (L) | 0.59 (L) | 0.74 (L) |
| TE4 | 0.22 (R) | 0.19 (R) | 4.81 (R) | 30.6 (R) | 0.58 (R) | 0.53 (R) |
| | 0.22 (L) | 0.25 (L) | 1.03 (L) | 0.32 (L) | 39.3 (L) | 0.29 (L) |
| MEC-Corr | 0.15 (R) | 0.15 (R) | 0.52 (R) | 0.30 (R) | 0.53 (R) | 0.51 (R) |
| | 0.18 (L) | 0.22 (L) | 0.49 (L) | 0.30 (L) | 0.46 (L) | 0.27 (L) |

*Abbreviations*: RMSE, root mean square error; L(R), the fiber-optic probe on the left (right); H., Heating; Cont., Continuous; Interm., Intermittent.

## Supporting Material

### In-vivo Canine MRgLITT Ablation

***Animal preparation and maintenance:*** Each canine was anesthetized using a subcutaneous injection of Atropine (0.05 mg/kg) and an intramuscular administration of Xylazine hydrochloride (1.5 mg/kg). A hole (approximately 3 x 3 mm$^2$) was drilled into the canine's skull, and the dura mater was carefully ruptured to place the laser applicator. The canine was positioned in a prone posture within the bore of the 3T MR scanner for imaging. Intermittent intravenous injections of Propofol (1-1.5 mg/kg) were administered to maintain anesthesia throughout the experiment.

***Pre-and post-ablation imaging:*** Before MRgLITT treatment, T1-weighted (see **Table S1**, 3D MPRAGE) MR images were first acquired to check the registration between the laser applicator and the temperature monitoring plane. Then the PRF temperature data were acquired using the multi-echo GRE sequence during the LITT procedure. Finally, post-interventional MR images were taken 20 minutes after the LITT treatment to assess the acute changes in the ablation zones. Gadolinium (0.2 ml/kg) contrast was injected, and MR images including T2-weighted images, fluid-attenuated inversion recovery (FLAIR), and T1-weighted images (i.e., 3D MPRAGE) were obtained using the scanning parameters listed in **Table S1**. Finally, the canines were euthanized.

**Table S1.** Acquisition parameters for the pre-and post-ablation MRI in in-vivo canine ablation experiments.

|  | 2D T2w | 2D FLAIR | 3D MPRAGE |
|---|---|---|---|
| TR (ms) | 2000 | 8000 | 7.6 |
| TE (ms) | 90 | 104 | 3.4 |
| TI (ms) | - | 2500 | - |
| FOV (mm$^2$) | 130 x 130 | 130 x 130 | 150 x 150 |
| Thickness (mm) | 4 | 4 | 0.7 |
| Sampling Matrix | 350 x 350 | 220 x 180 | 214 x 214 |
| Acquisition Time | 4 min | 2 min 48 s | 4 min 44 s |

*Abbreviation*s: TR, repetition time; TE, echo time; TI, inversion time; FOV, the field of view; MPRAGE, magnetization-prepared rapid gradient-echo; FLAIR, fluid-attenuated inversion recovery.

**Table S2.** The criteria of 5-point scale grading for temperature imaging.

| Score | Boundary Smoothness of Hotspots | Temperature Map Artifacts | Overall Thermometry Reliability |
|-------|--------------------------------|---------------------------|--------------------------------|
| 1 | Very rough boundary | Severe temperature artifacts | No ablation reliability |
| 2 | Rough boundary | Many deficits in temperature maps | Poor ablation reliability |
| 3 | Moderate boundary | Moderate temperature artifacts | Moderate ablation reliability |
| 4 | Smooth boundary | Few temperature artifacts | Good ablation reliability |
| 5 | Very smooth boundary | No visual artifacts | Excellent ablation reliability |

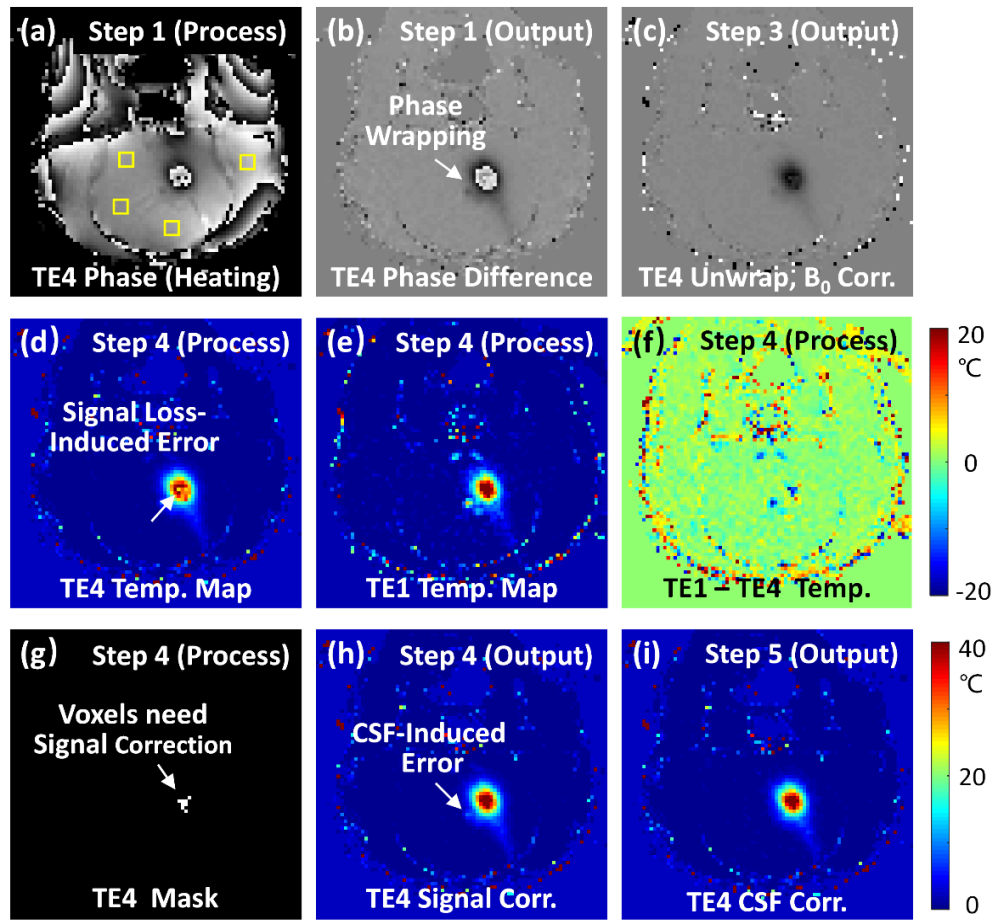

**Figure S1.** Intermediate images of the multi-echo-based temperature data processing algorithm. Each subfigure includes text in the upper right corner identifying the processing step it belongs to, as well as whether it represents an intermediate process or output of the step. (a) Example phase map of a time frame at TE4 = 19 ms obtained during heating in Step 1. Four thermally isolated ROIs applied for $B_0$ drift correction are indicated as yellow rectangles. (b) Phase difference map calculated by complex phase subtraction in Step 1 between (a) and the reference phase map (before heating). Phase wraps (white arrow) occur at the heating center where large temperature changes happen. (c) Phase-unwrapped and $B_0$ drift-corrected phase difference map achieved in Step 3. (d) Temperature map calculated from (c) in Step 4. The white arrow indicates the signal loss-induced temperature errors. (e) Temperature map calculated from data acquired at TE1 = 4 ms in Step 4 before CSF error correction. (f) The temperature difference between (e) and (d). (g) Erroneous voxels need signal artifact correction at TE4 = 19 ms in Step 4. Note that only voxels at the heating center are selected. (h) Signal loss-corrected temperature map obtained in Step 4. The white arrow indicates the residual CSF-induced temperature errors. (i) CSF-corrected temperature map achieved in Step 5. *Abbreviations*: Corr., correction; Temp., temperature.

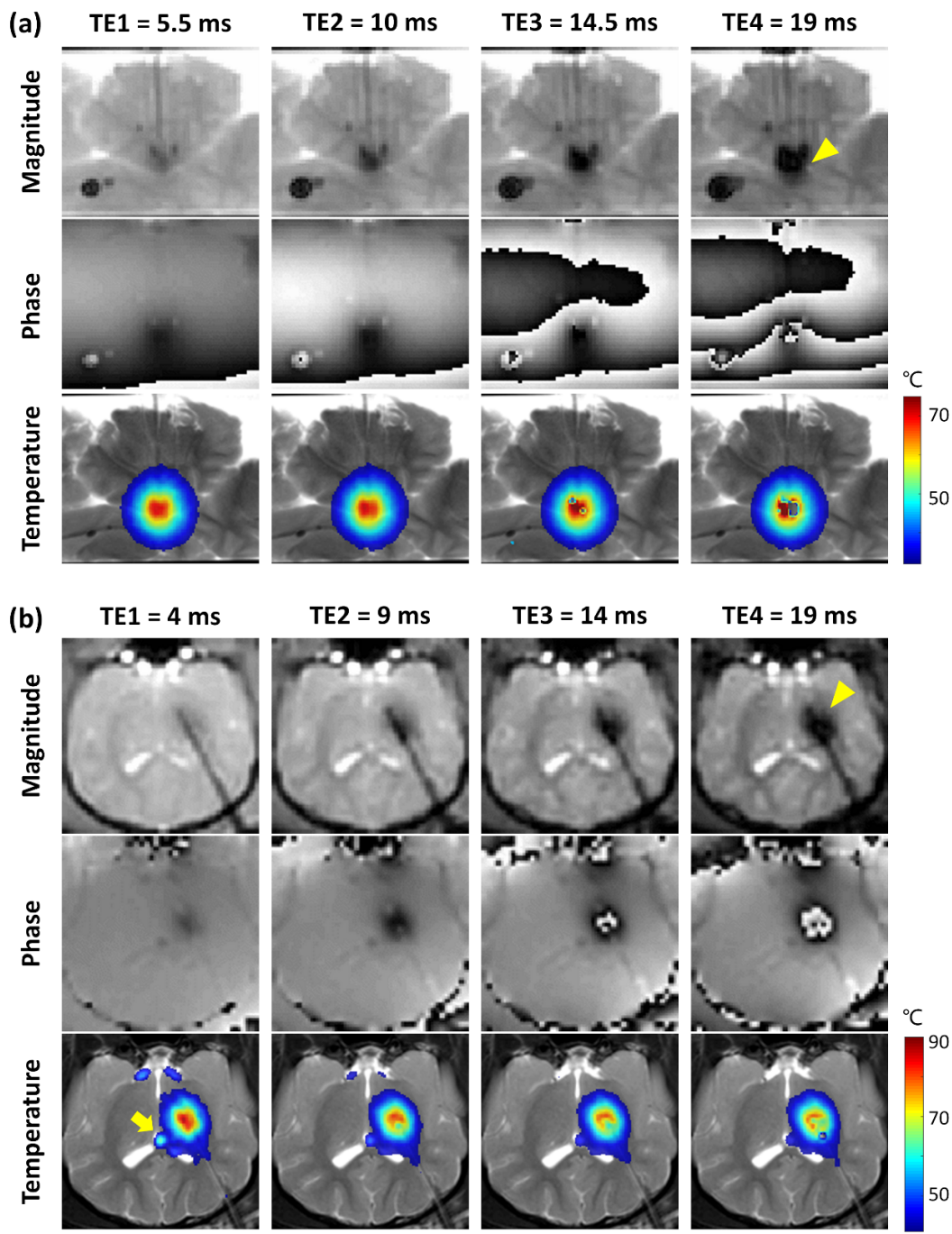

**Figure S2.** Zoomed-in magnitude, phase, and temperature images obtained from the ex-vivo swine brain (a) and in-vivo canine brain (b). The temperature maps are calculated from each TE using the traditional single-echo PRF algorithm. More phase wraps are present in the images with longer TEs, while the image contrast is also enhanced. Note that the intense temperature leads to signal loss in the magnitude images (pointed by the yellow arrowheads), and further translates to phase and temperature errors at the heating center. Note that CSF pulsation causes erroneous high temperatures within the ventricles, as pointed out by the yellow arrow in (b). The CSF-induced temperature errors are more apparent at shorter TEs. Also notice that TE1 in (b) demonstrates a rougher temperature boundary at the hotspot, indicating lower TNR.

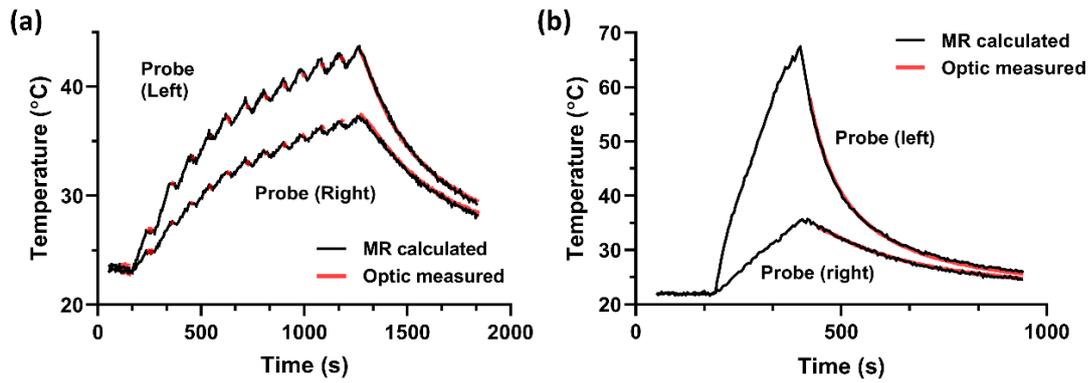

**Figure S3.** MR calculated temperatures (i.e., MEC-Corr results) within one single pixel near the fiber-optic probe (black line) and the fiber-optic-measured temperatures (red line) as a function of time. (a) Intermittent heating mode in the gel phantom experiment; (b) continuous heating mode in the ex-vivo pork experiment. Note that the fiber-optic measured temperatures are only presented at the temperature drop stages.

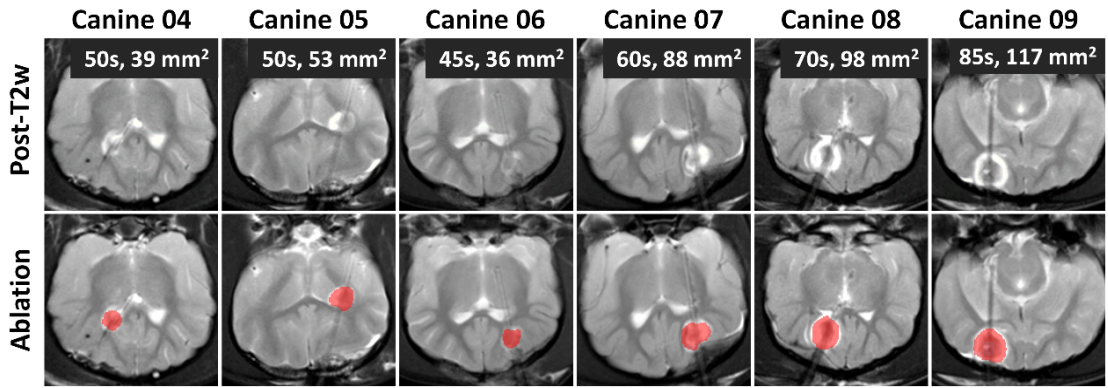

**Figure S4.** Thermal ablations of Canine 04-09 with different ablative laser doses (8W, tens of seconds) at different locations. The white contours on the post-ablation T2-weighted MR images (top row) match well with the ablation estimates of thermal damages calculated from the MEC-Corr temperature maps (bottom row). The upper right corner of each subfigure in the top row indicates the MRTI estimated ablation area, which varies from less than 40 mm$^2$ to around 120 mm$^2$ depending on the ablation location and the applied ablative dose.

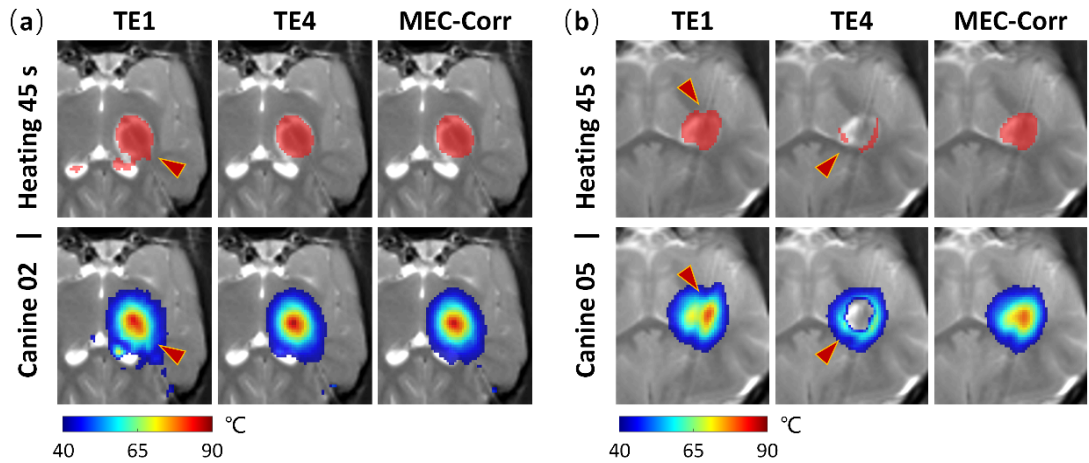

**Figure S5.** Thermal ablations of two example canines (i.e., Canine 02 and 05) estimated from TE1, TE4, and MEC-Corr temperature measurements, respectively. Canine 02 is unaffected by the MRTI signal voids, while Canine 05 is affected by the MRTI signal voids. The top row illustrates the MRTI-estimated ablation zones (shown in red) overlaid on the post-ablation T2-weighed images, while the bottom row displays the temperature maps acquired during intense heating. In (a), the ablation estimate from TE1 is less accurate due to CSF-induced temperature artifacts, whereas TE4 and MEC-Corr yield accurate estimations. In (b), signal loss-induced temperature artifacts affect the estimation of TE4 ablation, but MEC-Corr still manages to make a satisfactory prediction based on accurate temperature mapping. Furthermore, the ablation estimations from TE1 show rougher margins than those from TE4 or MEC-Corr in both (a) and (b) due to the rougher hotspot boundaries (i.e., lower TNR) in temperature imaging.

**Movie S1.** Zoomed-in magnitude (top row), phase (second row), and temperature maps (bottom row) of the in-vivo canine brain using the multi-echo GRE sequence (Frame 2-70). From left to right: TE1 = 4 ms, TE2 = 9 ms, TE3 = 14 ms, and TE4 = 19 ms. Temperature maps are calculated from each TE using the traditional single-echo PRF algorithm.

**Movie S2.** Temperature imaging during laser heating for (a) gel phantom (intermittent heating, Frame 50-300, Slice 1), (b) pork tissue (continuous heating, Frame 36-100, Slice 2), and (c) swine brain (intermittent heating, Frame 28-260, Slice 2). From left to right: 1) calculated from TE1 data via the conventional single-echo thermometry; 2) calculated from TE4 data via the conventional single-echo thermometry; 3) calculated from the 4-echo data using the proposed multi-echo thermometry. Note that the multi-echo algorithm (MEC-Corr) outperforms the single-echo thermometry (TE1 and TE4), demonstrating fewer signal loss-induced temperature artifacts, and improved TNR (i.e., smoother boundaries of the hotspot).

**Movie S3.** Temperature imaging of 3 example canines (i.e., Canine 01 - 03) under LITT heating from Frame 2-70 of Slice 2. From left to right: 1) calculated from TE1 data via the conventional single-echo thermometry; 2) calculated from TE4 data via the conventional single-echo thermometry; 3) calculated from the 4-echo data using the proposed multi-echo thermometry. Note that the multi-echo algorithm (MEC-Corr) outperforms the single-echo thermometry (TE1 and TE4), demonstrating fewer signal loss-induced or CSF-induced temperature artifacts.